# Self-assembly in solution of a reversible comb-shaped supramolecular polymer

*Sandrine Pensec[a], Nicolas Nouvel[a], Audrey Guilleman[a], Costantino Creton[b], François Boué[c], Laurent Bouteiller[a]*[*]

[a] UPMC Univ Paris 06, UMR 7610, Chimie des Polymères, F-75005 Paris, France, and CNRS, UMR 7610, Chimie des Polymères, F-75005 Paris, France

[b] Physico-chimie des Polymères et des Milieux Dispersés, UMR 7615, UPMC-CNRS-ESPCI, 10 rue Vauquelin 75231 Paris cedex 05, France

[c] Laboratoire Léon Brillouin, UMR 12 CNRS-CEA, 91191 Gif-sur-Yvette Cedex, France

[*] Corresponding author. E-mail: laurent.bouteiller@upmc.fr

ABSTRACT. We report a single step synthesis of a polyisobutene with a bis-urea moiety in the middle of the chain. In low polarity solvents, this polymer self-assembles by hydrogen bonding to form a comb-shaped polymer with a central hydrogen bonded backbone and polyisobutene arms. The comb backbone can be reversibly broken, and consequently, its length can be tuned by changing the solvent, the concentration or the temperature. Moreover, we demonstrate that the bulkiness of the arms has a strong influence on both the self-assembly pattern and the length of the backbone. Finally, the number of polyisobutene arms can be controlled, by simply mixing with a low molar mass bis-urea. This system thus combines a tunable structure and a dynamic backbone in solution. It is worth investigating its self-healing properties in bulk.

**Introduction**

Supramolecular polymers are chains of small molecules held together through reversible non-covalent interactions.[1-3] The dynamic character of such weak interactions is responsible for the appearance of new properties, as compared to those of usual covalent polymers. For example, these materials can display thermoreversible polymer-like properties (such as visco-elasticity), or even form self-healing elastomers.[4]

It is well known in the field of macromolecular science that the architecture of a polymer can have a significant effect on its rheological or mechanical properties. Therefore, there is a strong incentive to design and investigate the properties of supramolecular polymers with various architectures, such as macrocyclic,[5-10] star-shaped,[11-13] hyperbranched,[14] or reversibly cross-linked.[4,15-18] In this respect, comb-shaped supramolecular polymers[19] were among the first supramolecular polymers to be described and their original properties have been well recognized.[20,21] However, most of these comb-shaped supramolecular polymers consist of a covalent backbone decorated with side-chains that are reversibly linked to the backbone (Figure 1a); the reverse situation, where the backbone itself is dynamic (Figure 1b), has been very rarely reported.[22]

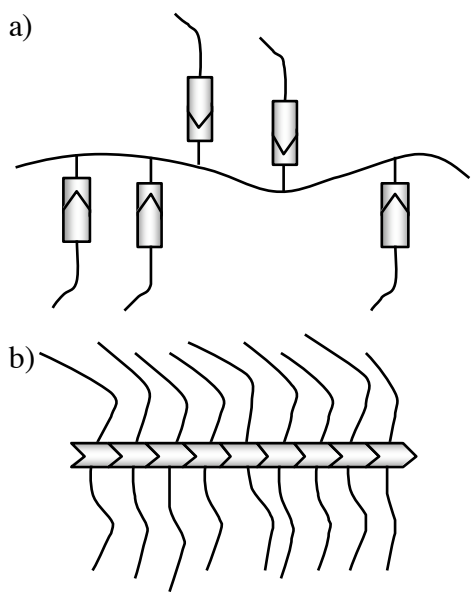

**Figure 1.** Schematic structure of a comb-shaped supramolecular polymer with a covalent (a) or a dynamic (b) backbone.

To build such a comb-shaped supramolecular polymer with a dynamic backbone, we chose the bis-urea synthon as the self-assembling unit, because of its strong self-association and its straightforward synthetic accessibility. Moreover, we have previously reported that bis-urea based low molar mass compound (**EHUT**, Figure 2) self-assembles in non-polar solvents into two distinct dynamic supramolecular polymer structures.[23] Depending on solvent, concentration and temperature, either long hydrogen bonded filaments with a single molecule in the cross-section, or even longer and more rigid tubes with three molecules in the cross-section are formed (Figure 3a). This competition between two different self-assembled structures opens the possibility to design responsive systems.

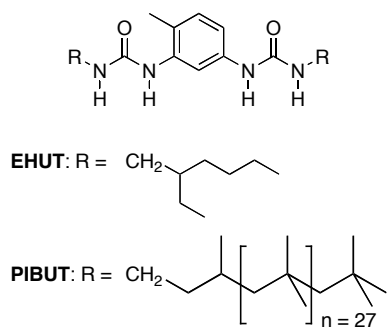

**Figure 2.** Structure of bis-ureas **EHUT** (EthylHexylUreidoToluene) and **PIBUT** (PolyIsoButeneUreidoToluene).

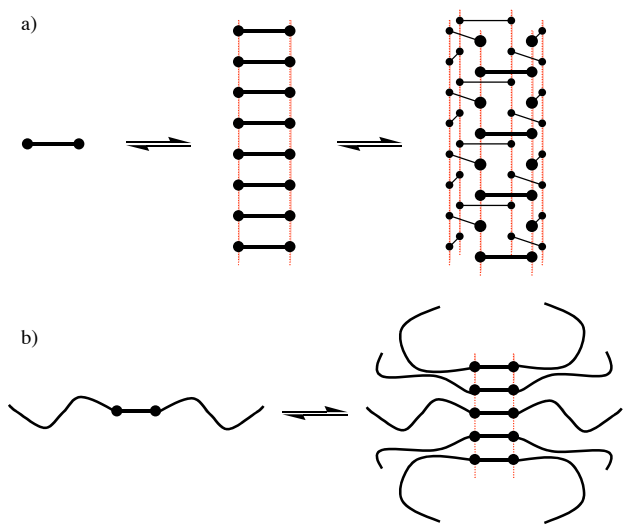

**Figure 3.** Schematic supramolecular arrangements for **EHUT** (a) and **PIBUT** (b). Hydrogen bonds are represented by red dotted lines connecting the urea functions. For more detailed **EHUT** models, see reference 24.

Polyisobutene, which has often been used in the context of hydrogen bonded supramolecular assemblies,[25-27] was chosen as the polymer side-chain because of its good solubility and absence of interfering hydrogen bonding groups. Therefore, we report in the present article the characterization of solutions of macromolecular bis-urea **PIBUT** (Figure 2).

**Experimental Section**

**Synthesis.** The synthesis of **EHUT** was described previously.[28] Non functional polyisobutene **PIB** ($M_n$ = 2800 g/mol, $M_w/M_n$ = 1.7) was obtained from Acros. Synthesis of 2,4-bis-(polyisobuteneureido)toluene **PIBUT**: 2,4-toluenediisocyanate (98% from Aldrich) (3.5 mL, 24.5 mmol) was added, at room temperature and under nitrogen to a stirred solution of amino-functional polyisobutene (**PIB-NH$_2$**: Kerocom PIBA, 60% solution in hydrocarbon, from BASF) (150 g) in dry THF (90 mL). After 24h, the reaction mixture was precipitated under vigorous stirring into 1L of ethyl acetate. A viscous oil decanted. After 24h, the upper phase was eliminated and the product was dried under vacuum for one month, to give a rubbery solid **PIBUT** (47.5 g). $^1$H NMR (200MHz, CDCl$_3$/d$_6$-DMSO (90/10 v/v)) see Supporting Information: δ (ppm) = 7.78 (s, 1H, Ar-N*H*), δ = 7.42 (s, 1H, Ar-*H*), δ = 7.12 (s, 1H, Ar-*H*), δ = 6.98 (s, 1H, Ar-*H*), δ = 6.80 (s, 1H, Ar-N*H*), δ = 5.80 (s, 1H, CH$_2$-N*H*), δ = 5.46 (s, 1H, CH$_2$-N*H*), δ = 2.99 (m, 4H, C*H*$_2$-NH), δ = 1.97 (s, 3H, Ar-C*H*$_3$), δ = 1.5-0.5 (m, 545H, C*H*$_2$-*C*H(*CH*$_3$)-C*H*$_2$ and (-C(*CH*$_3$)$_2$-*CH*$_2$)$_n$ and –C(*CH*$_3$)$_3$. $M_{n,NMR}$ = 3490 g/mol. $^{13}$C NMR (75MHz, CDCl$_3$/d$_6$-DMSO (90/10 v/v)): δ (ppm) =155.2 (*C*=O), δ = 137.7/137.0/129.5/119.5/112.1/110.7 (*Ar*), δ = 58.7 (-*C*(CH$_3$)$_2$-CH$_2$)$_n$), δ = 57.2 (CH$_2$-*C*H(CH$_3$)-CH$_2$), δ = 37.1 (-C(CH$_3$)$_2$-*C*H$_2$)$_n$), δ = 34.8 (N-*C*H$_2$-CH$_2$), δ = 31.7 (-*C*(CH$_3$)$_3$), δ = 31.6 (-C(*C*H$_3$)$_3$), δ = 30.4 (-C(*C*H$_3$)$_2$-CH$_2$)$_n$), δ = 25.7 (CH$_2$-CH(*C*H$_3$)-CH$_2$), δ = 22.0 (CH$_2$-CH(*C*H$_3$)-CH$_2$), δ = 16.6 (Ar-*C*H$_3$). SEC (THF, polystyrene calibration): $M_n$ = 2700 g/mol, $M_w/M_n$ = 1.2. MALDI-TOF (dithranol, Na$^+$): $M_{exp(n=9)}$ = 1493.24 g/mol, $M_{th(n=9)}$ = 1493.44 g/mol. DSC (2°C/min, N$_2$): $T_g$ = -73°C, $T_m$ = 68°C.

**Results and Discussion**

**1. Synthesis.** The bis-urea **PIBUT** was obtained by reacting an excess of amino-functional polyisobutene (**PIB-NH$_2$**) with 2,4-toluenediisocyanate. After purification by precipitation, the structure of the product was identified by $^1$H and $^{13}$C NMR spectroscopy. Size exclusion chromatography (SEC) showed a monomodal distribution with a low polydispersity index ($I_p$ = 1.2), proving that the excess of amino-functional polyisobutene had been washed off (Figure 4). The degree of polymerization was calculated from NMR signals, using the integration ratio between methylene protons of the repeat unit at 0.93 ppm and an aromatic proton at 7.42 ppm. The total degree of polymerization was found to be 54 (ie n = 27), corresponding to $M_n$ = 3490g/mol. The structure of **PIBUT** was also confirmed by the agreement between experimental molar masses measured by MALDI-TOF mass spectrometry and the theoretical molar mass.

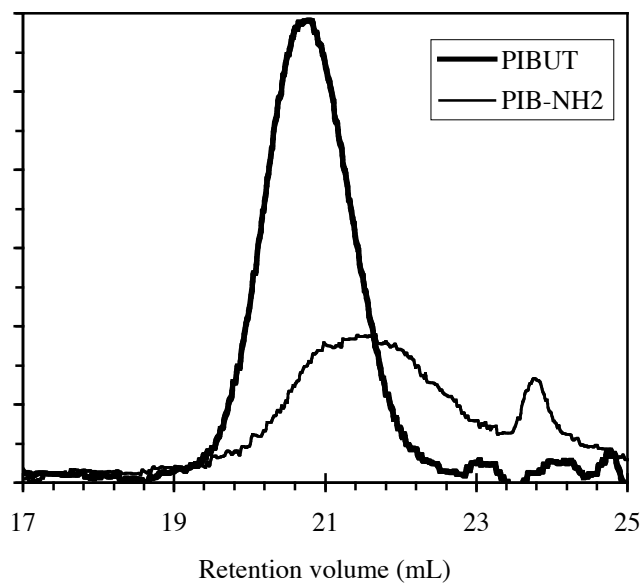

**Figure 4.** SEC trace for **PIBUT** and its amino-functional polyisobutene precursor **PIB-NH$_2$** (THF, refractive index detection).

**2. Viscosity of solutions.** Whether **PIBUT** self-assembles in solution can be qualitatively probed through the influence of solvent on the viscosity of **PIBUT** solutions. Figure 5 shows that the viscosity of **PIBUT** solutions increases significantly in the order tetrahydrofurane < chloroform < toluene < heptane. In contrast, solutions of **PIB** (a non hydrogen bonded polyisobutene of similar molar mass) have roughly the same viscosity, regardless of the solvent. This shows that the variation of viscosity for **PIBUT** solutions is not related to any potential difference in solvation of the polyisobutene arms, but rather to the influence of solvent on the strength of intermolecular hydrogen bond. The similar viscosity of **PIBUT** and **PIB** in tetrahydrofurane, means that **PIBUT** does not form any significant supramolecular assembly in this hydrogen bonding solvent. However, in a less competitive solvent such as chloroform, hydrogen bonds between urea functions can occur, as confirmed by FTIR spectroscopy (see below). Moreover, decreasing the polarity (from chloroform to toluene and to heptane) strengthens hydrogen bonds, allowing for further self-assembly and therefore increasing the viscosity.

Furthermore, it is of interest to compare the viscosity of **PIBUT** solutions to the viscosity of the low molar mass bis-urea **EHUT**. In tetrahydrofurane, where self-assembly is negligible, **PIBUT** is more viscous than **EHUT**, due to its one order of magnitude larger molar mass. In hydrocarbon solvents however, the ranking is reversed: **PIBUT** forms viscous solutions, whereas **EHUT** forms viscoelatic gels due to the entanglement of very long hydrogen bonded assemblies.[28-30] The comparatively much stronger effect of the solvent on **EHUT** viscosity than on **PIBUT** viscosity is an indication that the supramolecular assemblies formed are significantly different.

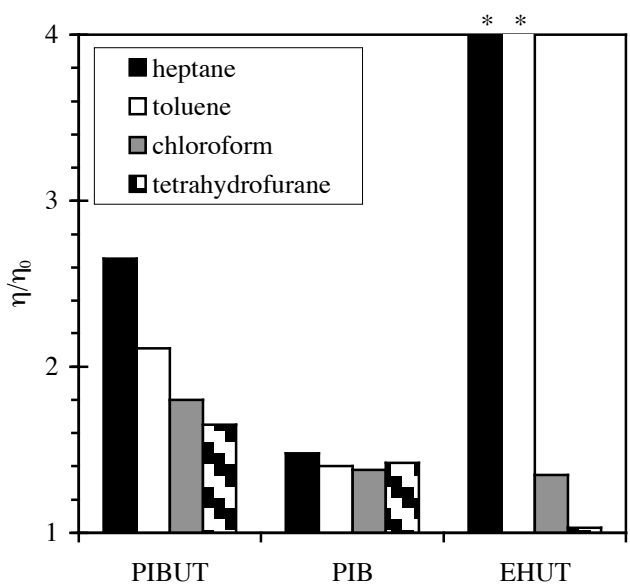

**Figure 5.** Relative viscosity ($\eta/\eta_0$) measured at 25 °C, for solutions of **EHUT** (24mM, 10g/L), **PIB** (24mM, 67g/L) and **PIBUT** (24mM, 83g/L) in several solvents. * The viscosity of **EHUT** solutions in heptane and toluene is much higher ($\eta/\eta_0 \gg 100$).[31]

**3. Characterization of the macromolecular structure.** The structure of the assembly formed by **PIBUT** was further characterized by SANS in $d_8$-toluene solution. Figure 6 shows that the scattered intensity reaches a plateau value at low q, which means that the scattering objects are of limited size. At high q, a $q^{-2}$ dependence (characteristic for Gaussian chains) is found, but in the intermediate q range, a decrease stronger than $q^{-2}$ is present and can be emphasized in a Kratky representation (inset of Figure 6). This maximum in Kratky representation is characteristic for branched structures. Therefore, a quantitative fit of the data was attempted with the form factor of a Gaussian comb.[32] Four parameters are necessary to describe such a comb: its molar mass (M), its radius of gyration (Rg), the number of arms per comb (f), and the fraction of chain segments in the backbone ($\lambda$). In the present case, the number of arms is directly linked to the comb molar mass, so that only three independent parameters were adjusted to fit the data (see details in Supporting Information). Figure 6 shows that an excellent fit is obtained with the following parameter values: M = 30000g/mol (and thus f = 17 arms per comb); Rg = 71Å; and $\lambda$ = 0.15. A fit was also attempted with the form factor for a Gaussian star, but the fit is not as good (see Figure S3 in Supporting Information). Considering the simplicity of the model used here, the good agreement with the data may seem surprising. Therefore, the influence of the polydispersity of the comb backbone on the scattered intensity was assessed. It is shown in Supporting Information that the effect of polydispersity is in fact very limited and does not qualitatively change the present results. Another concern is the use of Gaussian chain statistics: due to the dense packing of the arms, both the arms and the main-chain may be stretched out. However, the $q^{-2}$ dependence at high q unambiguously shows that at least a fraction of the chains follow Gaussian statistics. These may be free chains in solution (**PIBUT** monomers and very short combs) or the part of the arms farther from the backbone. If part of the arms are more stretched than the rest, they form a more compact layer. The corresponding scattering contribution would therefore be closer to a Porod scattering, varying like $q^{-4}$. This scattering would be hidden below the scattering of the Gaussian chains. But in fact, the fraction of chain segments involved in the compact layer is measured by $\lambda$. The low value deduced from the fit ($\lambda=0.15$) confirms the suitability of the model. As far as the backbone is concerned, its stretching is likely, but the difference between a rod shape and a Gaussian coil is pronounced only for values of DP significantly larger than the value found here ($DP_w=8.5$). In conclusion, at this concentration, **PIBUT** can be considered to form comb shaped supramolecular polymers in solution.[33]

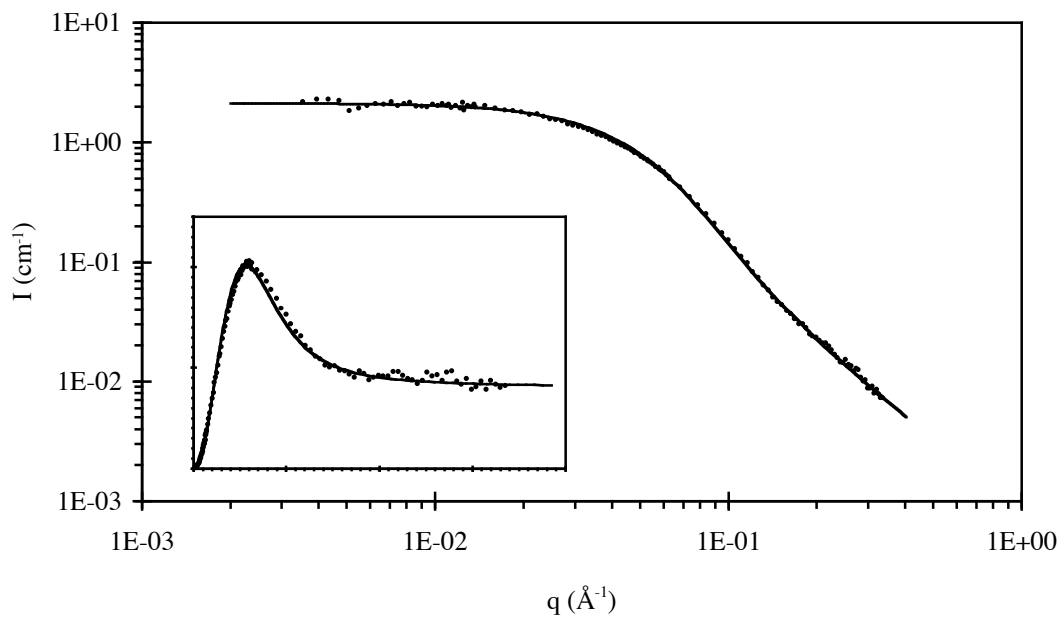

**Figure 6.** SANS intensity (I) versus scattering vector (q) for a solution of **PIBUT** in $d_8$-toluene at 11g/L (3.2mM) and 22°C. The inset shows the same data in a $q^2I$ versus q linear plot (which is much more sensitive to the scattering differences between various branched architectures, in particular with respect to the height and the width of the maximum, and the descending part on the right). The plain curve is a fit according to a model for comb-shaped polymers (see text and Supporting Information).

These SANS data give us the overall shape of the assemblies as well as the average size of one unit of the comb-like chain, but not the very local structure of the comb backbone. Based on previous data on **EHUT** and related bis-ureas,[23b,c] it is known that two different supramolecular arrangements can be envisaged: either hydrogen bonded filaments with a single molecule in the cross-section, or thicker and more rigid tubes with three molecules in the cross-section (Figure 3a). It was previously shown that FTIR spectroscopy can be used to discriminate between the two supramolecular structures,[23b] because the shape of the hydrogen bonded N-H vibration band is related to the exact hydrogen bonding pattern of the urea groups. Figure 7 shows the values of the ratio characterizing this band shape: a high value (c.a. 1.3) is attributed to the thick tubular structure, whereas a low value (c.a. 1.1) is attributed to the thin filament structure. Apparently, the supramolecular structure formed by **PIBUT** does not depend on the solvent nature: in all solvents tested, the same thin filament structure is obtained. This is in sharp contrast to the behavior of low molar mass bis-ureas such as **EHUT**, and implies that the thick tubular structure is unstable for **PIBUT**. This is probably due to a steric reason, because for a given backbone length, three times as many arms have to be accommodated in the tubular structure compared to the filament structure.

Therefore, a sensible model for the self-assembly of **PIBUT** in low polarity solvents is depicted in Figure 3b: comb-shaped objects are formed with a backbone made of a single filament of bis-urea moieties, and with polyisobutene arms.

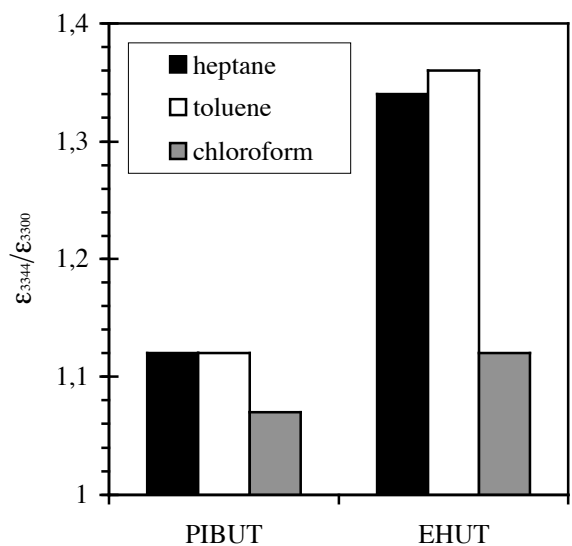

**Figure 7.** Ratio of FTIR absorbances at 3344 and 3300 cm$^{-1}$ for 12mM solutions of **EHUT** or **PIBUT** (25°C).

**4. Macromolecular effect on association strength.** Due to the reversibility of hydrogen bonds, the average length of these comb backbones can be expected to depend on parameters such as the solvent, the concentration and the temperature. Moreover, the steric bulk of the polyisobutene arms can be expected to be responsible for a weaker association than in the case of a low molar mass analog. The extent of this effect was studied by FTIR spectroscopy and isothermal titration calorimetry (ITC). Chloroform was chosen as the solvent, because in this solvent, both **EHUT** and **PIBUT** self-assemble into filaments with the same structure. Thus, any difference observed should be attributed to a difference in association strength. Figure 8 shows the FTIR spectra of **EHUT** and **PIBUT** at the same molar concentration in chloroform. In both cases, the hydrogen bonded N-H vibration band (3340-3280cm$^{-1}$) is the main band, but a weak free N-H vibration band (3450-3430cm$^{-1}$) can be detected. The intensity of this band is larger for **PIBUT** than for **EHUT** meaning that hydrogen bonding of the bis-urea moiety in **PIBUT** is indeed weaker than for **EHUT**. To quantitatively describe this effect over a large concentration range, it is useful to consider an association model describing the relative stability of monomer, dimer and all possible oligomers. The two-constant association model shown on Figure 9 was shown to adequately describe the assembly behavior of **EHUT** in chloroform.[31] Moreover, the association constants in this model were shown to be accessible through an ITC experiment, when a relatively concentrated solution is diluted, and the corresponding heat of dissociation is measured.[34] Figure 10 shows such an enthalpogram: the dissociation of **PIBUT** occurs over a much broader concentration range than the dissociation of **EHUT**, which means that the formation of **PIBUT** supramolecular polymers is much less cooperative. Both curves can be fitted by the association model of Figure 9.[35] The parameter values derived from the fit (Table 1) show that the dimerization step is not significantly affected by the bulk of the **PIBUT** arms, but that the subsequent steps are disfavored. This increased sensitivity to steric crowding for longer oligomers is not surprising and is the reason for the reduced cooperativity of **PIBUT** self-assembly. The knowledge of the association constants makes it possible to compute the molar mass of the comb-shaped supramolecular polymer over the whole concentration range (Figure S7).[31] Figure S7 shows that the formation of the **PIBUT** combs occurs only above 10$^{-3}$ mol/L in chloroform and that their growth is more sluggish than for **EHUT** supramolecular polymers. High molar masses are nevertheless reached at reasonable concentrations.

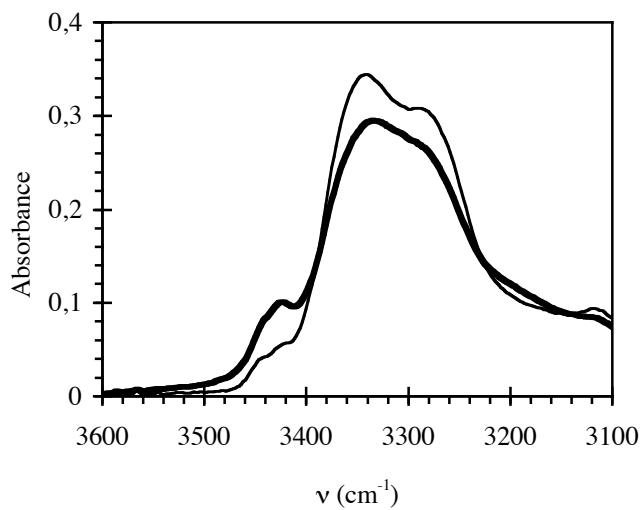

**Figure 8.** FTIR spectra for 12mM solutions of **EHUT** (plain) or **PIBUT** (bold) in chloroform (25°C).

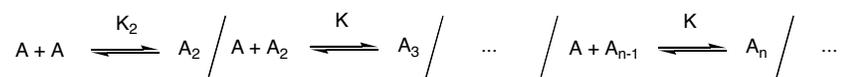

**Figure 9.** Association scheme describing the formation of a supramolecular polymer (A = monomer, $A_n$ = oligomer of degree of polymerization n).[3b]

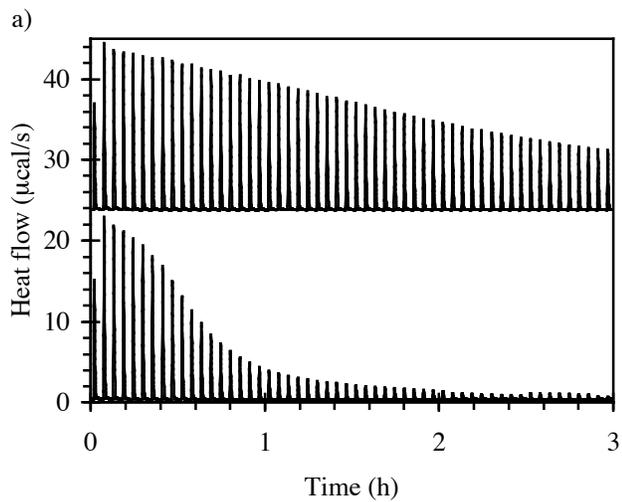

a)

b)

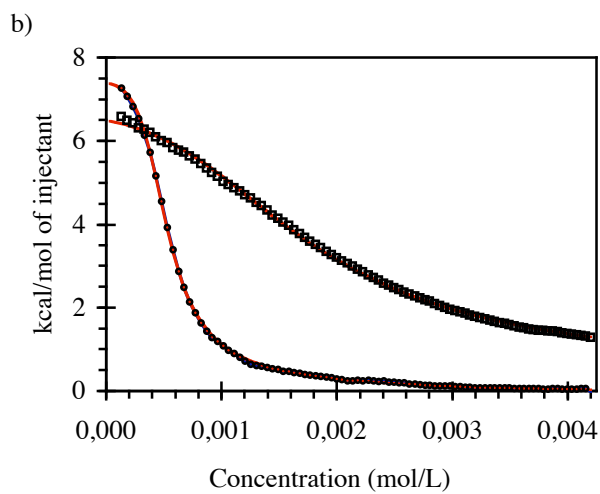

**Figure 10.** (a) Heat effect produced by injecting 3-µL aliquots of a 24mM chloroform solution of **PIBUT** (upper curve) or **EHUT** (lower curve) into chloroform (20°C). (b) Corresponding enthalpograms; the plain curves are the fits obtained with the model of Fig. 9 and the parameter values of Table 1.

**Table 1** Values for the parameters of the association scheme in Fig. 9, deduced from the ITC data of Fig. 10 (see also Fig. S5).

|  | $\Delta H_{assoc}$ (kJ/mol) | $K_2$ (L/mol) | $K$ (L/mol) |
|---|---|---|---|
| **EHUT** | -35 ± 4 | 58 ± 18 | 1700 ± 170 |
| **PIBUT** | -38 ± 4 | 63 ± 18 | 350 ± 40 |

**5. Comb-shaped copolymers.** The presence of the same associating bis-urea moiety in **PIBUT** and **EHUT** makes it potentially straightforward to form copolymers: simply mixing two solutions should afford a (probably statistical) copolymer.[36] In the present case, comb-shaped copolymers with an adjustable number of arms should be obtained. To check this possibility, heptane was chosen as solvent because the association is stronger than in the other more polar solvents. Moreover, the influence of temperature was monitored, because in heptane, **EHUT** self-assembles into filaments (above 75°C) or tubes (below 75°C), thus enabling to probe the possible copolymerization between **EHUT** and **PIBUT**, either in the filament or in the tube form. Figure 11 shows the result of variable temperature FTIR measurements on solutions of different compositions. The **EHUT** solution shows the expected transition between the low temperature tube and high temperature filament forms. The **PIBUT** solution shows no transition in the same temperature range, showing that only the filament form is stable, even down to –62°C. However, the equimolar mixture of the two solutions shows a transition temperature close to room temperature. To confirm this result and improve the precision of the transition temperature measurements, DSC experiments were performed on solutions of various compositions (Figure 12).[37] As a reference, the influence of concentration on the transition temperature of pure **EHUT** solutions is also plotted. The DSC results are in perfect agreement with the FTIR data. The fact that the transition temperature is affected by the composition proves that the two bis-ureas interact together and form some mixed assemblies, because if **EHUT** and **PIBUT** did not interact at all, then the transition temperature of **EHUT** would remain constant. Moreover, the fact that the transition temperature decreases when **PIBUT** is added, means that **PIBUT** interacts more favorably with the filament form of **EHUT** than with the tube form of **EHUT**. Thus, it seems reasonable to expect that comb-shaped copolymers with a thin filament backbone structure and an adjustable proportion of arms are formed at composition and temperature values corresponding to the region lying above the curve in Figure 12.

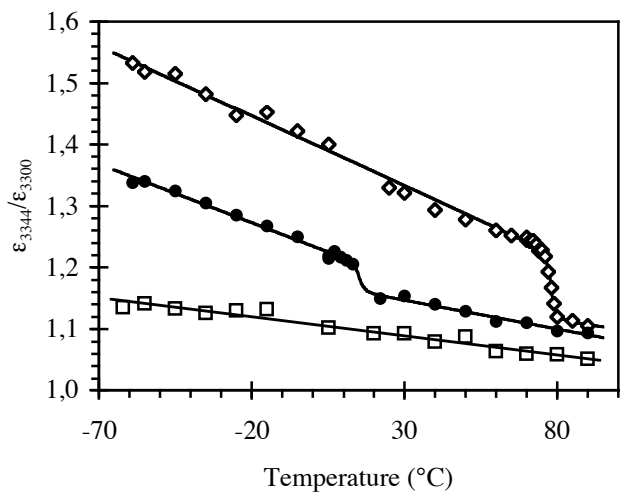

**Figure 11.** Ratio of absorbances at 3344 and 3300cm$^{-1}$ for 12mM heptane solutions of **EHUT** (◇), **PIBUT** (□), or their equimolar mixture (●).

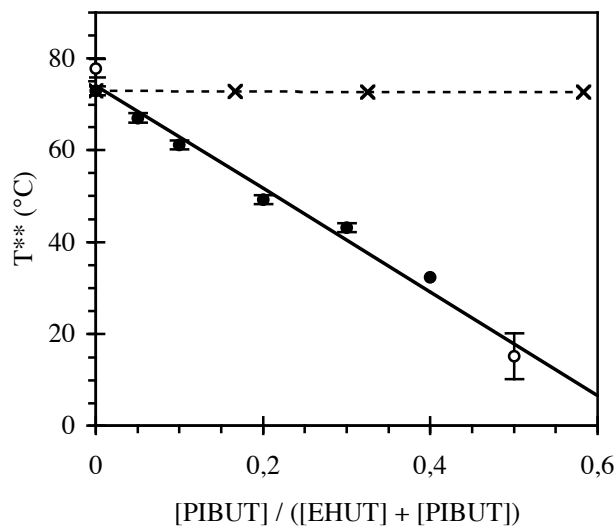

**Figure 12.** Transition temperature between tubes and filaments for **EHUT** / **PIBUT** mixtures, measured by FTIR (○) or DSC (●). (12mM solutions in heptane). The dotted line shows the evolution of the transition temperature for a pure **EHUT** solution at the same **EHUT** concentration as the mixtures.

**Conclusion**

We report the synthesis of a polyisobutene with a single bis-urea moiety in the middle of the chain. In low polarity solvents, this polymer self-assembles by hydrogen bonding to form a comb-shaped polymer with a central backbone, that can be reversibly broken. The length of the comb backbone can therefore be tuned by changing the solvent, the concentration or the temperature. Moreover, we demonstrate that the bulkiness of the arms has a strong influence on both the self-assembly pattern and the length of the backbone. Finally, the number of polyisobutene arms can be controlled, by simply mixing with a low molar mass bis-urea. This system thus combines a tunable structure and a dynamic backbone in solution. We are currently investigating the bulk properties of this new dynamic comb-shaped polymer.

**Acknowledgements.** We warmly thank A. Lange and BASF company for the gift of a Kerocom PIBA sample. C. Ng Pak Leung and L. Liejour are acknowledged for their contribution to this project.

**Supporting Information Available**. Experimental section for physicochemical characterizations and additional NMR and SANS data.

**References and Notes**

(1) Brunsveld, L.; Folmer, B. J. B.; Meijer, E. W.; Sijbesma, R. P. *Chem. Rev.* **2001**, 101, 4071.

(2) Ciferri, A. Supramolecular Polymers. Marcel Dekker Inc.: New York, 2005.

(3) (a) Binder, W. H.; Zirbs R. *Adv. Polym. Sci.* **2007**, 207, 1. (b) Bouteiller, L. *Adv. Polym. Sci.* **2007**, 207, 79.

(4) Cordier, P.; Tournilhac, F.; Soulie-Ziakovic, C.; Leibler, L. *Nature* **2008**, 451, 977.

(5) Ercolani, G.; Mandolini, L.; Mencarelli, P.; Roelens, S. *J. Am. Chem. Soc.* **1993**, 115, 3901.

(6) Yamaguchi, N.; Gibson, H. W. *Chem. Commun.* **1999**, 789.

(7) Abed, S.; Boileau, S.; Bouteiller, L. *Macromolecules* **2000**, 33, 8479.

(8) ten Cate, A. T.; Kooijman, H.; Spek, A. L.; Sijbesma, R. P.; Meijer, E. W. *J. Am. Chem. Soc.* **2004**, 126, 3801.

(9) Scherman, O. A.; Ligthart, G. B. W. L.; Sijbesma, R. P.; Meijer, E. W. *Angew. Chem. Int. Ed.* **2006**, 45, 2072.

(10) Ohkawa, H.; Takayama, A.; Nakajima, S.; Nishide, H. *Org. Lett.* **2006**, 8, 2225.

(11) Huang, F.; Nagvekar, D. S.; Slebodnik, C.; Gibson, H. W. *J. Am. Chem. Soc.* **2005**, 127, 484.

(12) Todd, E. M.; Zimmerman, S. C. *J. Am. Chem. Soc.* **2007**, 129, 14534.

(13) Bernard, J.; Lortie, F.; Fenet, B *Macromol. Rapid Commun.* **2009**, 30, 83.

manner between dimers and long oligomers. However, using a more complex model is not feasible because the data cannot yield reliable values for additional parameters. Therefore, $K_2$ and K have to be considered as apparent parameters, which make it possible to compare the behavior of **EHUT** and **PIBUT**.

# Self-assembly in solution of a reversible comb-shaped supramolecular polymer

*Sandrine Pensec, Nicolas Nouvel, Audrey Guilleman, Costantino Creton, François Boué, Laurent Bouteiller*

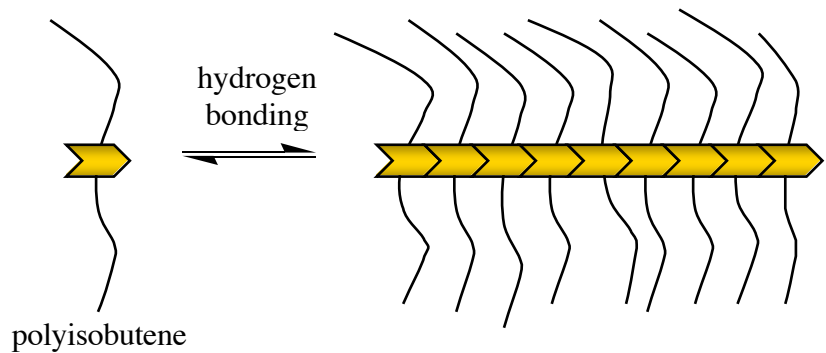